\documentclass{article}

\usepackage{PRIMEarxiv}

\usepackage{color}
\usepackage{graphicx}
\usepackage{algorithm}
\usepackage{algpseudocode}
\usepackage{bm}
\usepackage[utf8]{inputenc}
\usepackage{enumerate}
\usepackage{amsmath}
\usepackage{amssymb}
\usepackage{booktabs}
\usepackage{siunitx}
\usepackage{multirow}

\title{ModAn‑MulSupCon: Modality‑ and Anatomy‑Aware Multi‑Label Supervised Contrastive Pretraining for Medical Imaging}

\author{
  Eichi Takaya\\
  AI Lab \\
  Tohoku University Hospital \\
  Miyagi, Japan\\
  eichi.takaya.d5@tohoku.ac.jp \\
   \And
  Ryusei Inamori \\
  Department of Diagnostic Imaging \\
  Tohoku University Graduate School of Medicine \\
  Miyagi, Japan\\
   \\
}

\begin{document}
\maketitle

\begin{abstract}
Background and objective:
Expert annotations limit large‑scale supervised pretraining in medical imaging, while ubiquitous metadata (modality, anatomical region) remain underused. We introduce ModAn‑MulSupCon, a modality‑ and anatomy‑aware multi‑label supervised contrastive pretraining method that leverages such metadata to learn transferable representations.
\newline
Method:
Each image’s modality and anatomy are encoded as a multi‑hot vector. A ResNet‑18 encoder is pretrained on a mini subset of RadImageNet (miniRIN, 16,222 images) with a Jaccard‑weighted multi‑label supervised contrastive loss, and then evaluated by fine‑tuning and linear probing on three binary classification tasks—ACL tear (knee MRI), lesion malignancy (breast ultrasound), and nodule malignancy (thyroid ultrasound).
\newline
Result:
With fine‑tuning, ModAn‑MulSupCon achieved the best AUC on MRNet‑ACL (0.964) and Thyroid (0.763), surpassing all baselines ($p<0.05$), and ranked second on Breast (0.926) behind SimCLR (0.940; not significant). With the encoder frozen, SimCLR/ImageNet were superior, indicating that ModAn‑MulSupCon representations benefit most from task adaptation rather than linear separability.
\newline
Conclusion:
Encoding readily available modality/anatomy metadata as multi‑label targets provides a practical, scalable pretraining signal that improves downstream accuracy when fine‑tuning is feasible. ModAn‑MulSupCon is a strong initialization for label‑scarce clinical settings, whereas SimCLR/ImageNet remain preferable for frozen‑encoder deployments.
\end{abstract}


\section{Introduction}
In recent years, self‑supervised learning (SSL) has rapidly gained traction in medical imaging \cite{uelwer_survey_2025}\cite{rani_self-supervised_2024}. 
The primary reason is the substantial cost and effort required from clinical experts for annotation, which makes large‑scale supervised pretraining difficult to scale. 
Although hospital PACS archives contain vast amounts of unlabeled images, much of this data has remained unused. 
To bridge this gap, label‑free self‑supervised pretraining has attracted increasing attention. 

Prior work on SSL for medical imaging has explored a variety of approaches, including pretext tasks such as jigsaw puzzles \cite{sugawara_breast_2025}, contrastive‑learning methods \cite{wu_distributed_2022}, and masked‑autoencoder‑based techniques \cite{zhuang_mim_2025}. 
However, pretraining strategies that explicitly leverage large datasets spanning multiple modalities and anatomical regions remain underexplored. 
While large datasets for pretraining such as RadImageNet \cite{mei_radimagenet_2022} are becoming available, methods that exploit them have largely been limited to supervised learning with disease labels or simple MAE‑style objectives \cite{mei_radimagenet_2022}\cite{liu2024vis}.

In this study, we propose a pretraining method that exploits two medical‑image–specific yet easily obtainable attributes: modality and anatomical region. These attributes can be acquired for virtually any medical image without specialist annotation, making them highly scalable.

Our approach encodes each image’s modality and anatomical region as independent one‑hot vectors, concatenates them into a multi‑hot label, and performs pretraining with a distance‑aware loss using this target.
This enables the encoder to learn representations that reflect modality‑specific physics and anatomical structure, which in turn can improve both the efficiency of fine‑tuning and downstream accuracy.

\section{Related Works}
\subsection{Self‑Supervised Learning (SSL) in Medical Imaging}
Early SSL for medical imaging primarily relied on pretext tasks such as rotation prediction and jigsaw puzzles \cite{li_rotation-oriented_2021}\cite{sugawara_breast_2025}. Since 2020, however, contrastive learning has been introduced: adaptations of SimCLR \cite{chen2020simple} and MoCo \cite{he_momentum_2020} to medical images have demonstrated strong performance and data efficiency \cite{ren_ukssl_2023}\cite{sowrirajan_moco_2021}. 
These methods treat alternative slices from the same study or augmented views as positives, while maintaining large queues of negatives to learn robust representations. 

Multi‑modal contrastive approaches that couple images with text have also been actively studied.
Using report‑paired X‑rays, methods such as GLoRIA \cite{huang_gloria_2021} and BioViL \cite{boecking_making_2022} jointly optimize global and local image–text alignment and achieve strong transfer even with limited labels. 
More recently, masked image modeling (MIM) has come to the fore as a pretraining paradigm for Transformer‑based networks \cite{zhou_self_2023}\cite{zhuang_advancing_2025}\cite{xu_swin_2023}.
Because MIM can be trained with a single reconstruction objective, it is easy to implement, and has been extended to 3D volumes \cite{yu_unest_2023} and temporal sequences \cite{buess_video-ct_2024}, as well as combined with contrastive learning in multitask pretraining \cite{tang2022self}. 

Despite recent progress, SSL pretraining on large public datasets (e.g., RadImageNet) has mostly relied on disease‑label supervision \cite{mei_radimagenet_2022} or simple MAE baselines \cite{liu2024vis} and has not explicitly leveraged hierarchical metadata such as modality or body part.

\subsection{Learning with Multi‑Label Targets}
Multi‑label learning seeks to predict multiple labels for a single input. 
Classical approaches include problem‑transformation methods such as Binary Relevance \cite{tsoumakas_mining_2010} and Label Powerset \cite{boutell_learning_2004}, as well as algorithm‑adaptation methods exemplified by ML‑KNN \cite{zhang_ml-knn_2007} and Rank‑SVM \cite{elisseeff_kernel_2001}.  

With the rise of deep learning, many studies have modeled label dependencies within the network. For example, CNN–RNN architectures \cite{wang_cnn-rnn_2016} learn label co‑occurrence patterns. Spatial Regularization Networks \cite{zhu_learning_2017} represent inter‑label dependencies with attention maps. ML‑GCN \cite{chen_multi-label_2019} leverages graph convolution over label graphs to improve accuracy on benchmarks such as MS‑COCO. 
In extreme multi‑label text classification, tree‑based methods such as AttentionXML \cite{you_attentionxml_2019} achieve both speed and high accuracy on corpora like Amazon‑3M. 

Methods for hierarchical multi‑label prediction have also been proposed. 
However, many are restricted to single‑path trees \cite{zhang_use_2022}\cite{kokilepersaud_hex_2025}. 
As an alternative, representing labels as multi‑hot vectors has been shown to be effective \cite{li_hierarchical_2025}.

Multi‑label problem settings are common in medical imaging. In chest X‑rays (e.g., CheXpert \cite{irvin_chexpert_2019}, MIMIC-CXR \cite{johnson_mimic-cxr_2024}), multiple conditions such as pneumonia and pneumothorax must be predicted simultaneously. 
Recent state‑of‑the‑art models combine CNN backbones with GCNs or supervised contrastive losses to capture label co‑occurrence \cite{mao_imagegcn_2022}\cite{jaiswal_scalp-supervised_2021}\cite{chen_multi-label_2019}.

Our method represents the two meta-labels (modality and anatomy) as separate one-hot vectors and learns from their concatenation as a multi-hot target.
It is closest to Binary Relevance (it avoids class explosion as the number of labels grows), yet remains compatible with deep learning in that a single shared network learns the representation. Accordingly, our approach can be viewed as Binary Relevance extended with hierarchical metadata within a supervised contrastive framework.

\section{Method}

\begin{figure}[htbp]
\begin{center}
\includegraphics[width=.95\columnwidth]{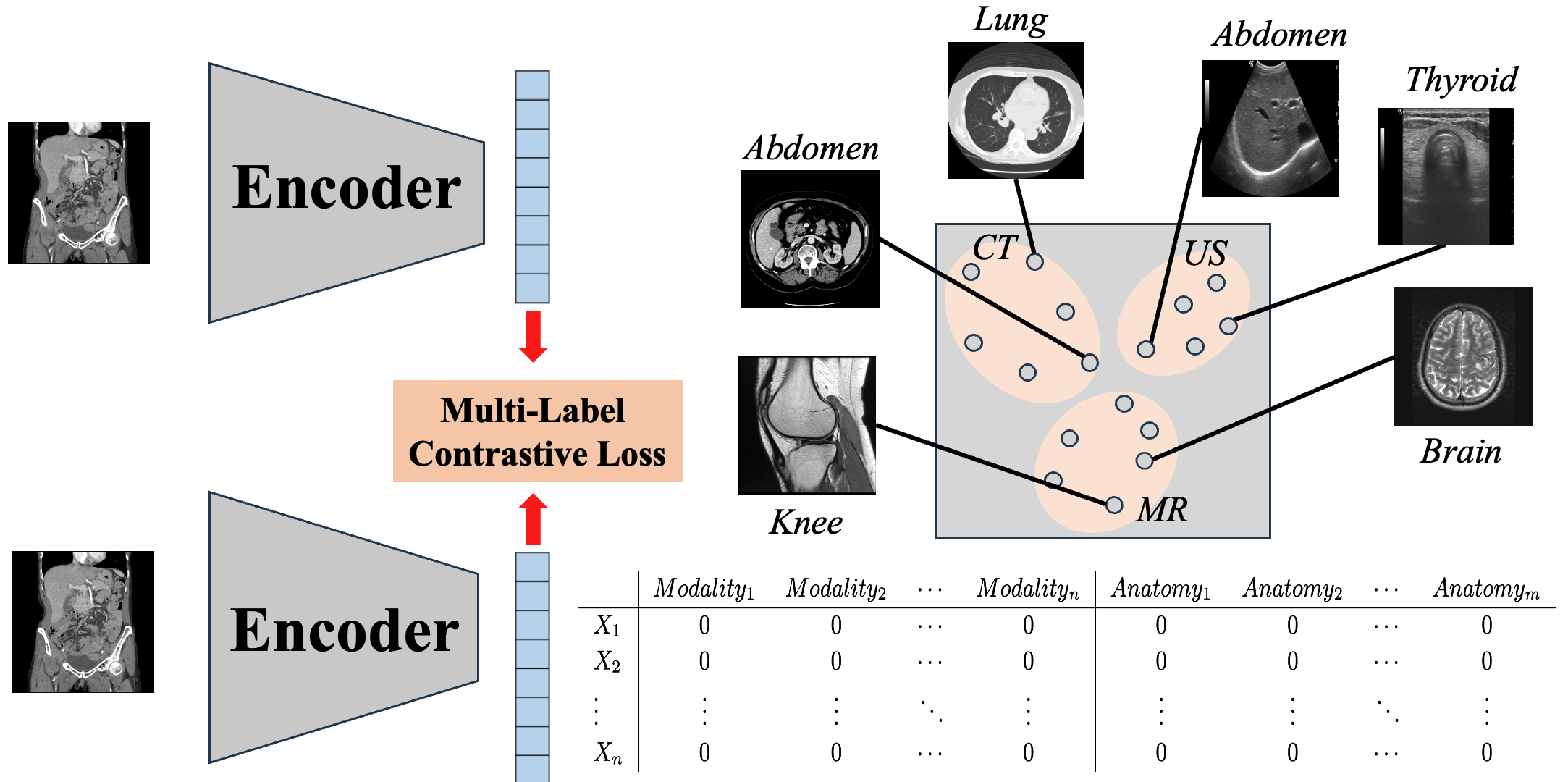}
\caption{ModAn‑MulSupCon: a shared encoder maps images to embeddings trained with a Jaccard‑weighted multi‑label supervised contrastive loss (Eq.~\ref{eq:supcon-ja}). Metadata are encoded as a multi‑hot vector (modality+anatomy), encouraging modality‑aware clusters (CT/US/MR) and anatomical coherence (lung, abdomen, thyroid, brain, knee).}
\label{overview}
\end{center}
\end{figure}

\subsection{Modality- and Anatomy-Aware Multi-Label Supervised Contrastive Learning (ModAn-MulSupCon)}
\label{sec:method}
Figure~\ref{overview} provides an overview of the proposed ModAn‑MulSupCon.
For an arbitrary collection of medical images:

\begin{enumerate}
  \item Map an input image $X$ with the encoder $f_{\theta}(\cdot)$
        to a $d$‑dimensional feature vector $\mathbf z$.
  \item Encode the image \emph{metadata}
        (modality and anatomical region)
        as a $k$‑dimensional multi‑hot label
        $\mathbf y\in\{0,1\}^{k}$.
  \item Train $f_{\theta}$ with a multi‑label contrastive loss.
\end{enumerate}

The pretrained encoder can subsequently be fine‑tuned for tasks such as
classification, detection, and segmentation.

\subsection{Multi‑Label Contrastive Loss}
To pretrain the encoder, we minimize a multi‑label supervised contrastive loss.

We first recall the InfoNCE loss for an anchor view $a$ and its positive view $p^\ast$ (two augmentations of the same image)~\cite{oord_representation_2018}:
\begin{equation}
  \mathcal{L}_{\mathrm{InfoNCE}}
  = -\log
    \frac{\exp\!\bigl(\mathbf z_a^{\top}\mathbf z_{p^\ast}/T\bigr)}
         {\displaystyle\sum_{k \in A(a)}
          \exp\!\bigl(\mathbf z_a^{\top}\mathbf z_k / T\bigr)} .
  \label{eq:step1}
\end{equation}
Here $\mathbf z_i \in \mathbb{R}^d$ denotes the $\ell_2$‑normalized representation of view $i$, $T>0$ is the temperature, and $A(a)$ is the set of all in‑batch views excluding the anchor, i.e., $A(a)=\{1,\ldots,2N\}\setminus\{a\}$ (two views per image in a batch of $N$ images).

If each image is associated with an explicit class label $y_i$,
the positive set for anchor $a$ is
\[
  P(a)=\{\,p \in A(a)\mid y_p=y_a\,\},
\]
and the supervised contrastive loss is~\cite{khosla2020supervised}
\begin{equation}
  \mathcal{L}_{\mathrm{SupCon}}
  = -\frac{1}{|P(a)|}
    \sum_{p\in P(a)}
    \log\frac{\exp\!\bigl(\mathbf z_a^{\top}\mathbf z_p/T\bigr)}
              {\displaystyle\sum_{k \in A(a)}
               \exp\!\bigl(\mathbf z_a^{\top}\mathbf z_k/T\bigr)} .
  \label{eq:step2}
\end{equation}

In the multi‑label setting, following prior work, each image $i$ is annotated with a binary vector $\mathbf y_i \in \{0,1\}^{k}$ indicating the presence of $k$ attributes (e.g., modality and anatomy). To account for partial matches between label sets, pairwise contributions are weighted by the Jaccard similarity~\cite{zaigrajew_contrastive_2022}:
\[
  w_{ap}
  = \mathrm{Jaccard}(\mathbf y_a,\mathbf y_p)
  = \frac{\lvert S(a)\cap S(p)\rvert}{\lvert S(a)\cup S(p)\rvert}
  \in [0,1],
\]
where $S(i)=\{\,m\mid (\mathbf y_i)_m=1\,\}$ is the set of active labels for sample $i$. Samples whose label similarity is at least a threshold $\tau \in [0,1]$ are treated as positives:
\[
  P_{\tau}(a)=\{\,p \in A(a)\mid w_{ap}\ge \tau\,\}.
\]
The resulting multi‑label supervised contrastive objective is
\begin{equation}
  \mathcal{L}_{\mathrm{MulSupCon}}
  = -\frac{1}{\lvert P_{\tau}(a)\rvert}
    \sum_{p\in P_{\tau}(a)}
    w_{ap}\,
    \log
    \frac{\exp\!\bigl(\mathbf z_a^{\top}\mathbf z_p/T\bigr)}
         {\displaystyle\sum_{k\in A(a)}
          \exp\!\bigl(\mathbf z_a^{\top}\mathbf z_k/T\bigr)} ,
  \label{eq:supcon-ja}
\end{equation}
which is known as the Multi‑Label Supervised Contrastive Loss~\cite{zaigrajew_contrastive_2022}.

\section{Experiments}
\subsection{Datasets}
\subsubsection{Pretraining Dataset}
\label{ssec:dataset-ja}
RadImageNet (RIN)~\cite{mei_radimagenet_2022} provides modality and anatomical metadata together with a large number of cases.
RIN contains approximately \num{1.3}~M medical images annotated with 165 disease and normal classes.

However, using the full dataset would require tens of GPU‑days per run, which is undesirable from the perspective of reproducibility and design exploration.
Therefore, we randomly sampled up to 100 images per class from the 165 classes in RIN (using all images for classes with fewer than 100 cases)
to construct the subset miniRIN. As a result, miniRIN comprises \num{16222} images in total.

\subsubsection{Downstream Dataset}
To evaluate the transferability of encoders trained on miniRIN, we selected three publicly available datasets with relatively few cases (Table~\ref{tab:down_ds}), each being a binary classification task.

\begin{table}[h]
    \centering
    \caption{Downstream datasets used for transfer‐learning evaluation.}
    \label{tab:down_ds}
    \begin{tabular}{lllll}
        \toprule
        Dataset & Modality & Task (binary) &  Images (class distribution) & Source \\
        \midrule
        Thyroid US & Ultrasound & Benign vs.~malignant nodule & 349 (288/61) & \cite{pedraza_open_2015} \\
        Breast US  & Ultrasound & Benign vs.~malignant lesion  & 779 (569/210) & \cite{al-dhabyani_dataset_2020} \\
        MRNet–ACL  & Knee MRI   & ACL tear vs.~intact ligament & 1021 (569/452) & \cite{bien_deep-learning-assisted_2018} \\
        \bottomrule
    \end{tabular}
\end{table}

\subsection{Experimental Setting}
In this study, we compared five initialization settings, with primary emphasis on classifiers fine‑tuned from the proposed ModAn‑MulSupCon pretraining model:
\begin{enumerate}
  \item ModAn‑MulSupCon (proposed)
  \item miniRIN SimCLR pretrained model (miniRINSimCLR)
  \item 165‑class supervised classifier trained on miniRIN (miniRIN165)
  \item ImageNet‑pretrained model (ImageNet)
  \item Random initialization (Scratch).
\end{enumerate}

For each setting, we performed both fine‑tuning (updating all layers) and linear probing (training only a linear classifier with the encoder frozen) on three downstream tasks, and evaluated performance using AUC.  
\paragraph{Data splits and repetitions.}
All methods were evaluated on the same fixed 5-fold partition for each downstream dataset. To characterize variance arising from training stochasticity - primarily mini-batch sampling - we repeated the entire 5-fold training/evaluation procedure ten times while keeping the data splits unchanged. In each repeat, models were trained/evaluated on all five folds and the fold-wise AUCs were averaged to obtain a single summary value. Thus, each method yielded ten summary AUCs, which were used for paired comparisons via the Wilcoxon signed-rank test (two-sided, $\alpha=0.05$).

\paragraph{Representation visualization.}
For qualitative analysis, we visualized learned representations from each encoder. Specifically, we extracted the penultimate feature vectors. These vectors were embedded into two dimensions using UMAP (Uniform Manifold Approximation and Projection) \cite{mcinnes2018umap}. To focus on pretraining effects, we visualized the miniRIN training set itself; for ImageNet and Scratch, features were computed on the same miniRIN training images for comparability, with no downstream or test images included.

\subsection{Implementation Details}
We used ResNet‑18 as the backbone and modified the first convolution to accept a single‑channel input (kernel $3{\times}3$, stride $1$), while keeping the remaining layers unchanged. During contrastive pretraining, we adopted a SimCLR‑style \texttt{TwoCropTransform}: each image is augmented into two views using \texttt{RandomResizedCrop} with scale $[0.2,\,1.0]$, \texttt{RandomHorizontalFlip}, \texttt{ColorJitter} (applied with probability $0.8$), and \texttt{RandomGrayscale} (applied with probability $0.2$).

Pretraining configurations for all initializations are summarized in Table~\ref{tab:pretrain}.
\begin{table*}[t]
  \centering
  \caption{Pretraining configurations across initialization settings. Common augmentations: \texttt{RandomResizedCrop}, \texttt{RandomHorizontalFlip}, \texttt{ColorJitter} ($p{=}0.8$), and \texttt{RandomGrayscale} ($p{=}0.2$).}
  \label{tab:pretrain}
  \small
  \begin{tabular}{lcccccc}
    \toprule
    \textbf{Method} & \textbf{Loss / Init.} & \textbf{Temp.\,$T$} & $\boldsymbol{\tau}$ &
    \textbf{Optimizer (lr, mom / wd)} & \textbf{Epochs} & \textbf{Batch} \\
    \midrule
    ModAn-MulSupCon  & Multi-label SupCon & 0.07 & 0.3 & SGD (0.05, 0.9 / $10^{-4}$) & 1000 & 256 \\
    miniRINSimCLR    & NT-Xent            & 0.10 & –   & SGD (0.05, 0.9 / $10^{-4}$) & 1000 & 256 \\
    miniRIN165       & Cross-entropy      & –    & –   & Adam (default)               & 1000 & 256 \\
    ImageNet         & Off-the-shelf      & –    & –   & –                             & –    & –   \\
    Scratch          & –                  & –    & –   & –                             & –    & –   \\
    \bottomrule
  \end{tabular}
\end{table*}
In brief, ModAn-MulSupCon uses a multi‑label supervised contrastive loss with temperature $T{=}0.07$ and threshold $\tau{=}0.3$, optimized by SGD (learning rate $0.05$, momentum $0.9$, weight decay $10^{-4}$) for $1000$ epochs with batch size $256$. The miniRINSimCLR baseline uses NT‑Xent with $T{=}0.10$ under the same optimizer and schedule, and the miniRIN165 supervised baseline is trained with cross‑entropy using Adam (default settings). ImageNet and Scratch correspond to off‑the‑shelf ImageNet weights and random initialization, respectively.

For downstream evaluation, we applied a single training recipe to both fine‑tuning (all layers trainable) and linear probing (encoder frozen and only a linear classifier trained). As shown in Table~\ref{tab:downstream}, we trained for $10$ epochs with batch size $32$ using Adam (learning rate $1{\times}10^{-4}$) and a StepLR scheduler (decay at epoch $5$ with factor $\gamma{=}0.1$).

All implementations were in PyTorch~2.1.2, and experiments were run on a single NVIDIA Quadro RTX~8000 (48\,GB) GPU.

\begin{table}[t]
  \centering
  \caption{Hyperparameters for downstream fine‑tuning and linear probing (common across tasks).}
  \label{tab:downstream}
  \small
  \begin{tabular}{lccc}
    \toprule
    \textbf{Epochs} & \textbf{Batch size} & \textbf{Optimizer} & \textbf{LR scheduler} \\
    \midrule
    10 & 32 & Adam ($1{\times}10^{-4}$) & StepLR$(5,\,\gamma{=}0.1)$ \\
    \bottomrule
  \end{tabular}\\[2pt]
  \footnotesize
\end{table}

\subsection{Results}
\label{ssec:result}

Under fine‑tuning, ModAn‑MulSupCon achieved the best AUC on both ACL and Thyroid (Table~\ref{tab:finetune_auc}), surpassing all baselines ($p<0.05$ in both tasks).
For Breast, miniRINSimCLR yielded a slightly higher AUC.

By contrast, with the linear probing, miniRINSimCLR and ImageNet performed best on ACL and Breast, while the ModAn-MulSupCon yielded the lowest scores (Table~\ref{tab:lp_auc}).
On Thyroid, miniRINSimCLR was best; ModAn‑MulSupCon was significantly better than Scratch only.

The wall‑clock per‑epoch times were approximately 10\,min/epoch for pretraining, 0.7\,sec/epoch for fine‑tuning, and 0.1\,sec/epoch for linear probing.

\begin{figure}[t]
\begin{center}
\includegraphics[width=.99\columnwidth]{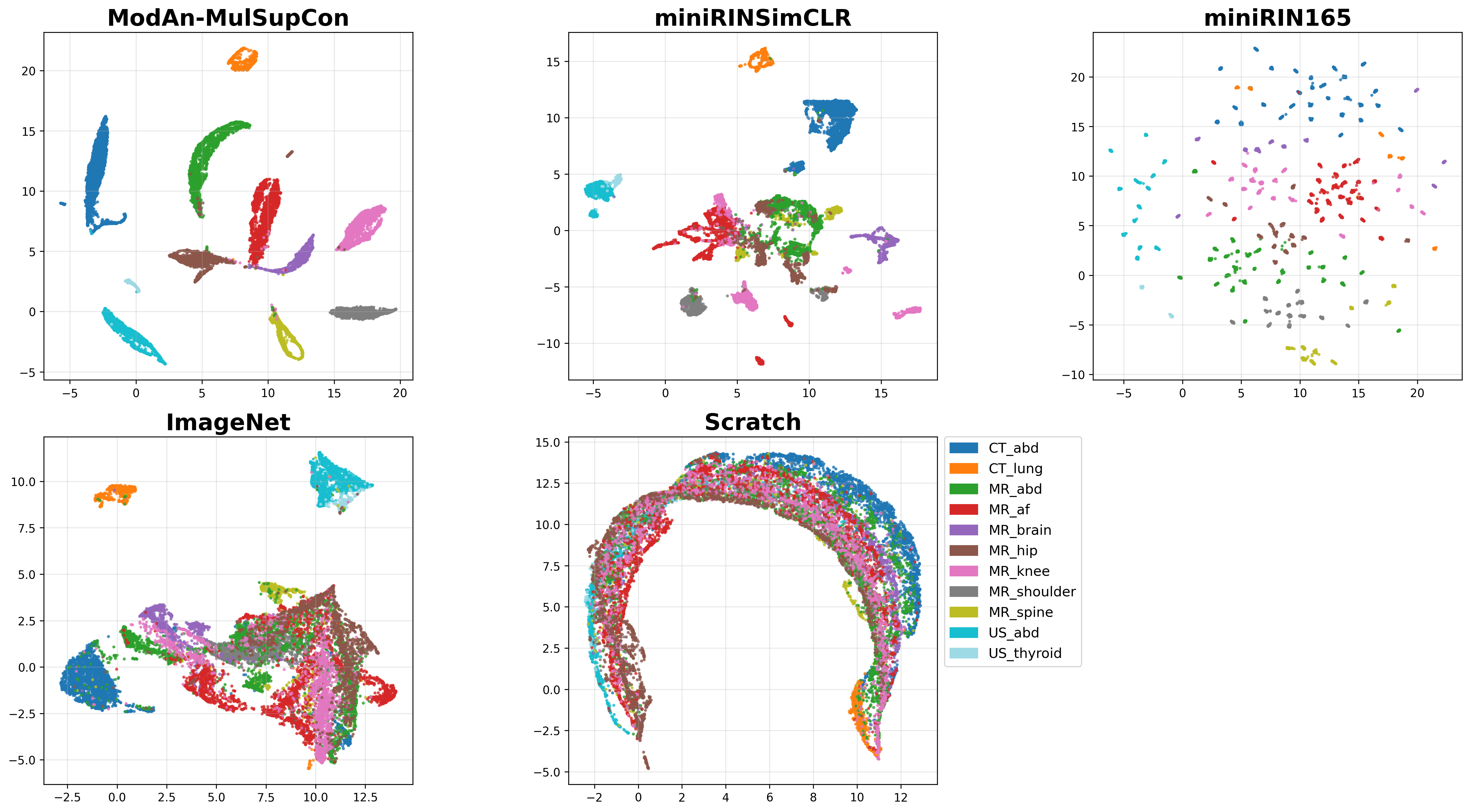}
\caption{UMAP projections of penultimate‑layer embeddings from five initializations— ModAn‑MulSupCon,  miniRIN‑SimCLR, ImageNet, miniRIN‑165, and Scratch. Each point is colored by the combined metadata (modality+anatomy): we assign an arbitrary but fixed color to each (modality, anatomy) pair and use the same mapping across panels; the colors are nominal only and do not encode similarity, density, or frequency.
}
\label{fig:umap}
\end{center}
\end{figure}

Figure~\ref{fig:umap} visualizes the learned embeddings with UMAP.
Each point is colored by the combined metadata (modality+anatomy): we assign an arbitrary but fixed color to each (modality, anatomy) pair and use the same mapping across panels; the colors are nominal only and do not encode similarity, density, or frequency.

\begin{table}[t]
  \centering
  \caption{Mean AUC and $p$‑values (Wilcoxon signed‑rank) after fine‑tuning with 10$\,\times\,$5‑fold CV, comparing each baseline to the proposed SSL.
  Best AUCs are \textbf{boldfaced},
  and $p\le 0.05$ is also \textbf{boldfaced}.}
  \label{tab:finetune_auc}
  \begin{tabular}{lcccccc}
    \toprule
    \multirow{2}{*}{Pretraining} &
      \multicolumn{2}{c}{ACL} &
      \multicolumn{2}{c}{Breast} &
      \multicolumn{2}{c}{Thyroid}\\
    & AUC & $p$-value & AUC & $p$-value & AUC & $p$-value\\
    \midrule
    ModAn-MulSupCon & {\bf 0.964} & -- & 0.926 & -- & {\bf 0.763} & --\\
    miniRINSimCLR       & 0.951 & {\bf$<$0.05} & {\bf0.940} & 1.000 & 0.685 & {\bf$<$0.05}\\
    miniRIN165  & 0.630 & {\bf$<$0.05} & 0.666 & {\bf$<$0.05} & 0.558 & {\bf$<$0.05}\\
    ImageNet     & 0.876 & {\bf$<$0.05} & 0.884 & {\bf$<$0.05} & 0.696 & {\bf$<$0.05}\\
    Scratch      & 0.907 & {\bf$<$0.05} & 0.884 & {\bf$<$0.05} & 0.677 & {\bf$<$0.05}\\
    
    \bottomrule
  \end{tabular}
\end{table}

\begin{table}[t]
  \centering
  \caption{Mean AUC and $p$-values for linear probing (pretrained encoder frozen; linear classifier only). Notation follows Table~\ref{tab:finetune_auc}.}
  \label{tab:lp_auc}
  \begin{tabular}{lcccccc}
    \toprule
    \multirow{2}{*}{Pretraining} &
      \multicolumn{2}{c}{ACL} &
      \multicolumn{2}{c}{Breast} &
      \multicolumn{2}{c}{Thyroid}\\
    & AUC & $p$-value & AUC & $p$-value & AUC & $p$-value\\
    \midrule
    ModAn-MulSupCon & 0.681 & -- & 0.717 & -- & 0.639 & --\\
    miniRINSimCLR       & {\bf0.844} & 1.000 & {\bf0.805} & 1.000 & {\bf0.657} & 0.935\\
    miniRIN165  & 0.796 & 1.000 & 0.723 & 0.986 & 0.630 & 0.188\\
    ImageNet     & 0.824 & 1.000 & 0.781 & 1.000 & 0.609 & 0.080\\
    Scratch      & 0.697 & 0.999 & 0.724 & 0.999 & 0.557 & {\bf$<$0.05}\\
    
    \bottomrule
  \end{tabular}
\end{table}

\section{Discussion}
We evaluated two regimes: full fine‑tuning of the pretrained encoder and linear probing with the encoder frozen. Under fine‑tuning, the ModAn‑MulSupCon achieved the best AUC on both ACL (0.964) and Thyroid (0.763), and ranked second on Breast (0.926; the best was miniRINSimCLR at 0.940) (Table~\ref{tab:finetune_auc}). By contrast, in linear probing, miniRINSimCLR and ImageNet were consistently superior, and the ModAn‑MulSupCon lagged behind on all three tasks (e.g., ACL: 0.681 vs.\ 0.844 for SimCLR; Table~\ref{tab:lp_auc}). These results suggest that the ModAn‑MulSupCon learns representations whose strength lies less in linear separability under freezing and more in their plasticity for task adaptation through fine‑tuning. In practice, it is therefore best viewed as an initialization tailored for scenarios that assume fine‑tuning.

ModAn‑MulSupCon supervises the encoder with a multi‑hot metadata target formed by concatenating modality and anatomy, and it weights pairwise terms by the Jaccard similarity (Eq.~\ref{eq:supcon-ja}, Fig.~\ref{overview}). This weighting treats different yet related meta‑attributes as “soft positives,” embedding hierarchical proximity into the representation space. As a result, the model learns non‑linear features aligned with modality‑specific physics and anatomical structure, which can be efficiently shaped into task boundaries during fine‑tuning. Notably, this strong inductive bias enabled competitive transfer even with only miniRIN (16{,}222 images) for pretraining, and on some tasks it surpassed ImageNet‑pretrained baselines.

As a complement, UMAP visualizations (Figure~\ref{fig:umap}) show that both the ModAn‑MulSupCon and SimCLR form clusters roughly organized by modality and anatomy.
Whereas SimCLR exhibits higher linear separability when the encoder is frozen, the ModAn‑MulSupCon yields more entangled clusters that nonetheless translate into superior accuracy after fine‑tuning. This pattern is consistent with the fine‑tuning advantage and frozen‑encoder disadvantage observed in Tables~\ref{tab:finetune_auc} and \ref{tab:lp_auc}.

On Breast (ultrasound), SimCLR achieved a slightly higher AUC (0.940) than the ModAn‑MulSupCon (0.926) (Table~\ref{tab:finetune_auc}). Ultrasound is strongly affected by device variability, speckle noise, and acquisition protocols; thus, coarse metadata such as modality and anatomy may be insufficient to constrain differences arising from subtle peri‑lesional textures or ROI selection. In such cases, instance‑level relational learning (e.g., SimCLR) can be advantageous for both linear separability under freezing and certain fine‑tuned settings. This observation indicates that a metadata‑driven inductive bias is not universally optimal; domains with high acquisition heterogeneity may benefit from finer‑grained metadata (e.g., device or probe type) or from hybridizing metadata supervision with instance contrastive learning.

From an operational standpoint, SimCLR or ImageNet is preferable when the encoder must remain frozen for lightweight deployment. When full fine‑tuning is feasible, the ModAn‑MulSupCon is more likely to deliver superior final performance. A practical guideline, therefore, is: use SimCLR/ImageNet for frozen encoders, and adopt the ModAn‑MulSupCon when fine‑tuning is possible.

This study has several limitations. First, the scalability assessment is limited to the miniRIN scale. Second, the downstream datasets offer restricted modality diversity, leaving untested regions. Third, our evaluation assumes a ResNet‑18 backbone; extensions to ViT‑style architectures and 3D backbones are needed.

\section{Conclusion}
We introduced ModAn‑MulSupCon, a modality‑ and anatomy‑aware multi‑label supervised contrastive pretraining approach that turns readily available metadata into a useful supervision signal. In transfer, it performs best when the encoder can be fully fine‑tuned, whereas generic pretraining such as SimCLR or ImageNet is preferable when the encoder must remain frozen. In practice, we recommend ModAn‑MulSupCon for scenarios that permit fine‑tuning and generic pretraining for deployments that require a frozen encoder.
Because modality and anatomy labels are broadly available without expert annotation, the proposed strategy offers an immediately deployable, label‑efficient pretraining signal for clinical pipelines operating under limited annotations.

Looking ahead, we will scale pretraining beyond miniRIN, enrich the metadata used for supervision with factors such as device or probe type, view, and laterality, integrate metadata supervision with instance‑level contrastive objectives, evaluate additional backbones including Vision Transformers and three‑dimensional encoders, and assess robustness across institutions and modalities while expanding to broader clinical classification settings.
\section*{Acknowledgments}
This work was supported by JSPS KAKENHI Grant Number JP24K15174.
\bibliographystyle{unsrt}  
\bibliography{modan}

\end{document}